\documentclass[a4paper,10pt]{amsart}

\newtheorem{theorem}{Theorem}

\begin{document}

\title{Holonomy groups and spacetimes}

\author{G.S. Hall and D.P. Lonie}

\address{Department of Mathematical Sciences\\
University of Aberdeen, Aberdeen AB24 3UE, Scotland, UK\\
e-mail: g.hall@maths.abdn.ac.uk}

\date{}

\begin{abstract}
A study is made of the possible holonomy group types of a
space-time for which the energy-momentum tensor corresponds to a
null or non-null electromagnetic field, a perfect fluid or a
massive scalar field. The case of an Einstein space is also
included. The techniques developed are also applied to vacuum and
conformally flat space-times and contrasted with already known
results in these two cases. Examples are given.
\end{abstract}

\maketitle

\section{Introduction}










In some recent papers attempts have been made to describe
space-times according to their holonomy group. The general
situation was discussed in the review [1] and the solution for a
vacuum and a conformally flat space-time was given, respectively
in [2] and [3]. In this paper both null and non-null
Einstein-Maxwell fields together with perfect fluids and massive
scalar fields will be similarly studied and the holonomy
classification for them obtained. The techniques developed will
also be used to give an alternative approach to the vacuum and
conformally flat cases.

There are several ways of classifying space-times. Two common ones
are through the Petrov type of the Weyl tensor (the Petrov
classification) and through the algebraic type of the energy-
momentum tensor (the Segre classification). For a review see e.g.
[4]. These are, however, pointwise in the sense that the Petrov or
Segre type may vary from point to point (subject to continuity
requirements which can be described topologically [5]). Thus such
classification systems, when imposed on the space-time as a whole,
require not only a restriction as to the algebraic type, but also
that type to be imposed globally (so that each point of the
space-time is of the "same type"). In practice these are really
{\it {local}} classifications (but rather useful nonetheless). An
alternative classification which is global in the sense that it
applies to the whole (or at least to the appropriate part) of the
space-time can be achieved by assuming the existence of certain
families (usually Lie algebras) of vector fields on the space-time
and which describe certain symmetries of it. This is again useful
(see, e.g. [4] [5]) but complicated because of the diverse nature
of the symmetries involved. The classification through holonomy
groups considered here, whilst incomplete and also suffering
itself from certain global {\it{types}} being assumed,
nevertheless considers a global property of the space-time
connection, namely the holonomy type.

Let $M$ be a space-time with $M$ and all structures on $M$ assumed
smooth. A standard notation will be used with a comma and a
semi-colon denoting the usual partial and covariant derivative and
with round and square brackets representing the usual
symmetrisation and skew- symmetrisation. Let $g$ be the space-time
metric with signature $ \left( { -  +  +  + } \right)$ with
associated Levi-Civita connection $\Gamma $ and curvature operator
$R'$ with tensor components $R^a _{\ bcd} $. For any $p \in M$ let
$C^k \left( p \right)$ denote the set of all piecewise $C^k \left(
{1\leq k\leq\infty } \right)$ closed curves at $p$. For each $c
\in C^k \left( p \right)$ there is an obvious vector space
isomorphism $f_c :T_p M \to T_p M$ (where $T_p M$ is the tangent
space to $M$ at $p$) defined by parallel transport around $c$ and
the set of all such maps arising from all members of $C^k \left( p
\right)$ is a group under the usual operations $f_{c_1 }  \circ
f_{c_2 }  = f_{c_1  \cdot c_2 } $ and $f_c ^{ - 1}  = f_{c^{ - 1}
} $ for $ c,c_1 ,c_2  \in C^k \left( p \right)$ . This group is
Lie-isomorphic to a Lie subgroup of the Lie group $ GL\left(
{4,{\bf R}} \right)$ and is independent of the degree of
differentiability $k$ and (up to isomorphism) of $ p \in M$ . This
group is abstractly denoted by $ \Phi$ and called the {\it {
holonomy group}} of $M$. If one repeats the above construction
this time restricting to members of $ C^k \left( p \right)$ which
are continuously or smoothly homotopic to zero (it matters not
which) one arrives at the {\it {restricted holonomy group}} of $M$
denoted abstractly by $ \Phi ^0$ . In fact $ \Phi ^0$ is the
identity component Lie subgroup of $\Phi$ . If $M$ is simply
connected $ \Phi ^0  = \Phi$ . (For further details of holonomy
theory including the results of this section, [6] is an excellent
text).

Throughout this paper $M$ will be assumed simply connected. It
then follows that $ \Phi ^0  = \Phi$ and hence that $ \Phi$ is a
connected Lie subgroup of the identity component $L_0$ of the
Lorentz group $L$. Let $\phi$ be the holonomy algebra (the Lie
algebra of $\Phi$ or $\Phi^{0}$ ) and $A$ the Lorentz algebra (the
Lie algebra of $L$ or $L_{0}$). Then $\phi$ is a subalgebra of $A$
and since $\Phi$ is now connected (since $M$ is simply connected)
the possibilities for $\Phi$ correspond in one-to-one fashion to
the subalgebras of $A$. The subalgebra structure for $A$ is well
known and given in table 1.

\begin{table}
\caption{The first column gives the labelling (following [8]) for
each subalgebra of $A$ (omitting the trivial flat case $R_{1}$ and
the case $R_{5}$ which is impossible for space-times [1]). The
second column gives a bivector basis for the subalgebra in terms
of a real null tetrad $ \left( {l,n,x,y} \right)$
 for all types except $R_{13}$ where a psuedo-orthonormal basis $
\left( {u,x,y,z} \right)$ is used. Here $ \sqrt 2 u = l - n,\sqrt
2 z = l + n$ and the only non-vanishing inner products are $ l^a
n_a  = x^a x_a  = y^a y_a  = z^a z_a  =  - u^a u_a  = 1$ . The
wedge product is used to describe bivectors so that, for example,
$ l \wedge n = 2l_{[a} n_{b]}$ . In the $R_{12}$ row $ 0 \ne e \in
{\bf R}$ . The third column lists the recurrent vector fields (up
to an obvious scaling) which cannot be globally scaled so as to be
covariantly constant. The fourth column lists in the $< >$
brackets a spanning set for the vector space of covariantly
constant vector fields on $M$. The final column lists the
subalgebra dimension. }

\begin{center}
\begin{tabular}{|l|l|l|l|l|}
\hline {\bf Type} & {\bf Lie Algebra} & {\bf Recurrent} & {\bf
Constant} & {\bf Dimension}
\\ \hline $R_{2}$ & $l\wedge n$ & $l, n$ & $<x, y>$ & $1$
\\ \hline $R_{3}$ & $l\wedge x$ & $-$ & $<l, y>$ & $1$
\\ \hline $R_{4}$ & $x\wedge y$ & $-$ & $<l, n>$ & $1$
\\ \hline $R_{6}$ & $l\wedge n, l\wedge x$ & $l$ & $<y>$ & $2$
\\ \hline $R_{7}$ & $l\wedge n, x\wedge y$ & $l, n$ & $-$ & $2$
\\ \hline $R_{8}$ & $l\wedge x, l\wedge y$ & $-$ & $<l>$ & $2$
\\ \hline $R_{9}$ & $l\wedge n, l\wedge x, l\wedge y$ & $l$ & $-$ &
$3$
\\ \hline $R_{10}$ & $l\wedge n, l\wedge x, n\wedge x$ & $-$ & $<y>$ &
$3$
\\ \hline $R_{11}$ & $l\wedge x, l\wedge y, x\wedge y$ & $-$ & $<l>$ &
$3$
\\ \hline $R_{12}$ & $l\wedge x, l\wedge y, l\wedge n+e(x\wedge y)$ & $l$
& $-$ & $3$
\\ \hline $R_{13}$ & $x\wedge y, y\wedge z, x\wedge z$ & $-$ & $<u>$ &
$3$
\\ \hline $R_{14}$ & $l\wedge n, l\wedge x, l\wedge y, x\wedge y$ & $l$
& $-$ & $4$
\\ \hline $R_{15}$ & $A$ & - & $-$ & $6$
\\ \hline
\end{tabular}
\end{center}
\end{table}

Now let $ p \in M$ and, in some coordinate system about $p$,
consider the set of matrices of the form
\begin{equation}
  R^a _{\ bcd} X^c Y^d ,R^a
_{\ bcd;e} X^c Y^d Z^e , \cdots
\end{equation}
where $ X,Y,Z \in T_p M$ . This set can be shown to constitute a
Lie algebra under matrix commutation which (up to isomorphism) is
independent of the coordinates chosen. It is denoted by $ \phi
'_p$ , referred to as the {\it {infinitesimal holonomy algebra}}
of $M$ at $p$, and is a subalgebra of both $\phi$ and $A$ for each
$ p \in M$ . The corresponding unique connected Lie subgroup
arising from $ \phi '_p$ is referred to as the {\it {infinitesimal
holonomy group}} of $M$ at $p$ and denoted by $ \Phi '_p$ . If  $
\dim \Phi '_p $ is independent of $p$ then for $ p,q \in M$ $ \Phi
'_p$ and $ \Phi '_q$ are (up to isomorphism) equal to each other
and also to $\Phi^0$ (and, since $M$ is simply connected, also to
$ \Phi$ ).

An important result in holonomy theory is the {\it {Ambrose-Singer
theorem}} (see e.g. [6]). Let $ p \in M$ and c be a differentiable
curve from $p$ to some point $ q \in M$ . Denoting parallel
transport along c from $p$ to $q$ by $\tau$ then if $ u,v \in T_p
M$ and $ R'$ is the curvature operator one can construct a linear
map $T_p M \to T_p M$ by
\begin{equation}
 w \to \tau ^{ - 1} \left[ {R'\left( {\tau
(u)\tau (v),\tau (w)} \right)} \right]
\end{equation}
for $ w \in T_p M$ . With $p$ fixed, and for all choices of $ u,v
\in T_p M$ , $ q \in M$ and $c$, the set of all such linear maps
spans (a representation of) the holonomy algebra $\phi$ of $M$.

This theorem essentially says that a representation of $\phi$ at $
p \in M$ can be obtained by choosing some $ q \in M$ , finding all
bivectors in the range of the curvature tensor at $q$ (i.e. all
bivectors of the form $ R^a _{\ bcd} H^{cd}$ at $q$) and parallely
transporting them along some curve $c$ to $p$. The bivectors which
occur at $p$ for all choices of $ q \in M$ and $c$ span $\phi$.

It is remarked here that a detailed study of the space-time {\it {
infinitesimal holonomy group}} has been given [7] following
earlier work in [8] and [9]. However these studies rely upon
certain (explicit or implicit) assumptions being made about the
constancy of the dimension of the infinitesimal holonomy group
over $M$. In this paper no such assumptions are made and interest
is focussed on the "full"{\it {holonomy group}} $\Phi$ of $M$.

\section{Holonomy Reducibility}
Let $ p \in M$ and, with the notation of the previous section,
define $\Phi _p$ by
\begin{equation}
  \Phi _p  = \left\{ {f_c :c {\rm \ a \ piecewise \
differentiable \ closed \ curve \ at \ }p} \right\}
\end{equation}
Then the holonomy group $\Phi$ of $M$ is called {\it {reducible}}
if for some (and hence any) $ p \in M$ and for some non-trivial
proper subspace $ V \subseteq T_p M$ , $V$ is invariant under each
member of $ \Phi _p$ . Otherwise $\Phi$ is called {\it {
irreducible}}. Such a subspace $V$ is called {\it {holonomy
invariant}}. Further, $\Phi$ is called {\it {non-degenerately
reducible}} if a (non-trivial proper) non-null holonomy invariant
subspace of $ T_p M$ exists at some (and hence every) $ p \in M$ .
If $\Phi$ is reducible, but not non-degenerately reducible, it is
called {\it {degenerately reducible}} [10].

Regarding the Lorentz group $ L_0$ as a 6-dimensional Lie group of
$4\times 4$ non-singular matrices preserving the Lorentz matrix $
\eta _{ab} = {\rm diag}( - 1,1,1,1)$ one may represent the Lorentz
algebra $A$ as the vector space of $4\times 4$ skew-symmetric
matrices (bivectors) in the usual way (see table 1). The holonomy
algebra $\Phi$ will then be viewed as a subalgebra of $A$ in this
representation. One then has the exponential map $ \exp :\phi  \to
\Phi$ defined as the usual exponential of matrices (see e.g.[11]).

A global, nowhere zero vector field $k$ on $M$ is called {\it {
recurrent}} if there exists a global covector field $q$ on $M$
such that in any coordinate domain $ k^a _{;b}  = k^a q_b $. One
can now collect together the following results (see e.g. [1] [12])

\begin{theorem}
Let $M$ be a simply connected space-time with (connected) holonomy
group $\Phi$ and holonomy algebra $\phi$. Then
\begin{enumerate}
\item $\Phi$ is reducible if and only if the members of $\phi$
admit a common eigenvector. \item $\Phi$ is reducible if and only
if $M$ admits a global recurrent vector field. \item The members
of $\phi$ admit a common eigenvector with zero eigenvalue if and
only if $M$ admits a non-zero global covariantly constant vector
field. \item The holonomy group $\Phi$ is reducible if and only if
it is not of type $R_{15}$.
\end{enumerate}
\end{theorem}

It is remarked here that a nowhere zero covariantly constant
vector field is recurrent but not necessarily conversely. In fact,
if $k$ is a recurrent vector field on $M$ then it gives rise in an
obvious way to a 1-dimensional holonomy invariant subspace at each
$ p \in M$ and $k$ is either everywhere timelike, everywhere
spacelike or everywhere null. Also if $k$ is recurrent, so is
$\alpha k$ for a nowhere zero $ \alpha :M \to {\bf R}$ . Then if
$k$ is recurrent and not null, the associated normalized (to $ \pm
1$) vector field obtained from $k$ is covariantly constant. This
result may fail if $k$ is null in the sense that it may not be
possible to globally scale $k$ so that it is covariantly constant.

Finally it is mentioned that the Lorentz group $ L_0$ has the nice
property that it is an exponential group, that is, the exponential
map $ \exp :A \to L_0$ is surjective (onto). Thus every member of
$ L_0$ is the exponential of some member of $A$ (see, e.g.[13]).
This property can fail for the (connected) Lie subgroups of $ L_0$
. However one does have the following property of any connected
Lie group $G$ (and, in particular for the connected Lie subgroups
of $ L_0$ ) that any member of $G$ is a product of finitely many
members of $G$ each of which is the exponential of a member of the
Lie algebra of $G$.

It is convenient at this point to collect together some results
that will be of use later. Let $ R_{ab}  = R^c _{\ acb}$ be the
components of the Ricci tensor of $M$, $ R = R_{ab} g^{ab}$ the
Ricci scalar and $ C^a _{\ bcd}$ the Weyl tensor components. Then
the curvature tensor can be decomposed as
\begin{eqnarray}
 R_{abcd}  & = C_{abcd}  + R_{a[c} g_{d]b}  + g_{a[c} R_{d]b}
- {\textstyle{1 \over 3}}RG_{abcd} \\
   & = C_{abcd}  + E_{abcd}  + {\textstyle{1 \over 6}}RG_{abcd}
   \nonumber
\end{eqnarray}
where
\begin{equation}
E_{abcd}  = \tilde R_{a[c} g_{d]b}  + \tilde R_{b[d} g_{d]a}
\end{equation}
\begin{equation}
G_{abcd}  = g_{a[c} g_{d]b}, \hspace{1cm} \tilde R_{ab}  = R_{ab}
- {\textstyle{1 \over 4}}Rg_{ab}  = E^c _{acb}
\end{equation}
The statement that $M$ is an Einstein space (so that $ R_{ab}  =
{\textstyle{1 \over 4}}Rg_{ab}$ ) is equivalent to either of the
statements $ \tilde R_{ab}  = 0$ or $ E_{abcd}  = 0$ .

The Einstein field equations are
\begin{equation} R_{ab}  -
{\textstyle{1 \over 2}}Rg_{ab}  = \kappa T_{ab}
\end{equation}
where $ T_{ab}$ are the components of the energy-momentum tensor
and $ \kappa  \ne 0$ is the gravitational constant. The following
duality properties are also useful (where * denotes the usual
duality operator)
\begin{equation} {}^ * C_{abcd}  = C_{abcd}^
*, \ \ {}^ * G_{abcd} = G_{abcd}^ *, \ \ {}^ * E_{abcd}  =  -
E_{abcd}^ *
\end{equation}
It follows that, when considered in the usual $6\times 6$
formulation as linear maps on bivector space at any $ p \in M$ ,
the maps arising from the tensors $ C_{abcd}$ , $ G_{abcd}$ and $
E_{abcd}$ have even rank. For the Weyl tensor at $p$ the rank is 2
in the Petrov type $N$ case, 4 for type $III$, 6 for types $II$
and $D$ and 4 or 6 for type $I$.

\section{Vacuum Space-Times}
This special case has been dealt with elsewhere [2] and the
following theorem proved

\begin{theorem}
Let $M$ be a simply connected, not flat, vacuum space-time. Then
the holonomy group $\Phi$ of $M$ is one of the types $R_{8}$,
$R_{14}$ or $R_{15}$.
\end{theorem}
The proof in [2] essentially consists of using the first equation
in (8), applied now to the curvature tensor in vacuum, to
establish that whenever a bivector belongs to the holonomy algebra
$\phi$ (in the sense of table 1) then so also does its dual. From
this it follows that the holonomy algebra is even-dimensional and
the Ambrose-Singer theorem together with a consideration of the
infinitesimal holonomy completes the proof.

A mild variation of this proof starts by assuming that $M$ is not
flat (otherwise $\Phi$ is of type $R_{1}$) and then supposes that
$\Phi$ is not of type $R_{15}$ and is hence reducible from theorem
1. The same theorem then shows that $M$ admits a recurrent vector
field $k$. If $k$ is covariantly constant on $M$ the Ricci
identity on $k$ gives $ R^a _{\ bcd} k^d  = 0$ from which it
follows that the curvature tensor (equal to the Weyl tensor in
vacuo) is of Petrov type $N$ (and hence of rank 2) and that $k$
spans the repeated principal null direction at those points of $M$
where the curvature is non-zero. Since $M$ is not flat it follows
that $k$ is a null vector field on $M$ and the curvature tensor is
constructed out of a dual pair of null bivectors with principal
null direction $k$ at those points where it is not zero. Since $k$
is covariantly constant and hence preserved under parallel
translation, the Ambrose-Singer theorem shows that $\phi$ consists
of a dual pair of null bivectors with principal null direction $k$
and hence that $\Phi$ is of type $R_{8}$. If $k$ is recurrent and
$M$ admits no covariantly constant vector fields then $k$ must be
null and $ R^a _{\ bcd} k^d$ cannot be identically zero on $M$.
The latter statement follows because otherwise the previous
argument using the Ambrose-Singer theorem would again lead to
$\Phi$ being of type $R_{8}$ and from table 1 and theorem 1(iii)
one then achieves the contradiction that $k$ is covariantly
constant on $M$. Thus if $ k_{a;b}  = k_a p_b$ for some covector
field $p$ on $M$ the Ricci identity on $k$ shows that on $M$ $
R_{abcd} k^d  = k_c F_{ab}$ where the bivector F is not
identically zero on $M$ and, where non-zero, is null with
principal null direction $k$ (from the vacuum condition and the
identity $ R_{[abc]d}  = 0$) . Hence at such a point $p$ the
Petrov type is $III$ with repeated principal null direction $k$
and, as is well known, the curvature tensor is constructed from
four independent bivectors at $p$. Thus $ \dim \phi '_p \geq 4$
and from table 1 and the fact that $\Phi$ is reducible $\Phi$ must
be of type $R_{14}$. This completes the proof.

If $\Phi$ is of type $R_{8}$ the Petrov type is $N$ at those
points where the curvature is non-zero. If $\Phi$ is of type
$R_{14}$ the Petrov type is $N$ where the curvature tensor is not
zero and where $ R_{abcd} k^d  = 0$ and type $III$ where $
R_{abcd} k^d  \ne 0$ . The vacuum pp-waves [14] are essentially
the only examples of holonomy type $R_{8}$. Examples of type
$R_{14}$ can be found in [15] [16]. The generic vacuum metric is
of type $R_{15}$ (e.g. the Schwarzschild metric).

\section{Einstein Space(-Times)}
In view of theorem 2, $M$ will be assumed a proper Einstein space
with $ E_{abcd}  \equiv 0$ and $ R = {\rm constant} \ne 0$ in (4).
Then (4) and (8) show that $ {}^ * R_{abcd}  = R_{abcd}^ *$ and
the argument in [2] again shows that $\phi$ is dual invariant in
the sense described at the beginning of the proof of theorem 2 and
hence that $ \dim \Phi$ is even. Table 1 then shows that the
possibilities for the type of $\Phi$ are $R_{7}$, $R_{8}$ $R_{14}$
and $R_{15}$. However if $\Phi$ were of type $R_{8}$ the curvature
tensor would be constructed entirely out of a dual pair of null
bivectors at each $ p \in M$ with principal null direction $ k \in
T_p M$ . But then, at $p$, $ R^a _{\ bcd} k^d  = 0$ $ \left( {
\Rightarrow R_{ab} k^b  = 0} \right)$ and so one finds the
contradiction that $ R = 0$ at $p$ (and, incidentally, that $M$
admits no covariantly constant vector fields). Thus one has the
following result.

\begin{theorem}
Let $M$ be a simply connected proper Einstein space(-time). Then
the holonomy group $\Phi$ of $M$ is one of the types $R_{7}$,
$R_{14}$ or $R_{15}$.
\end{theorem}
The recurrence condition and Ricci identity on any null recurrent
vector field $k$ on $M$ gives
\begin{equation}
k_a R^a _{\ bcd}  = k_a C^a _{\ bcd} + {\textstyle{1 \over
{12}}}R\left( {k_c g_{bd}  - k_d g_{bc} } \right) = 2k_b p_{[c;d]}
\end{equation}
Thus $ k_{[e} R^a _{\ b]cd} k_a  = 0$ and then
\begin{eqnarray}
\left( {\rm i}\right) \ \  k_{[e} C^a _{\ b]cd} k_a  =
{\textstyle{1 \over {12}}}R\left( {k_{[e} g_{b]c} k_d  - k_{[e}
g_{b]d} k_c } \right) \\ \left( {\rm ii}\right) \ \  k_{\ [e} C^a
_{b]cd} k^c k_a = 0 \nonumber
\end{eqnarray}
where the expressions in (i) are nowhere zero on $M$. Now if
$\Phi$ is of type $R_{7}$ it is known that there are two recurrent
(null) vector fields satisfying (10) (see table 1 or [12]) and
hence, using the Bel criteria for the Petrov types, such a
space-time is of Petrov type $D$ everywhere with these recurrent
null directions as the principal null directions. If $\Phi$ is of
type $R_{14}$ only one such recurrent (null) vector field exists
(table 1) and so the Petrov type is $II$ or $D$ at each point with
the recurrent null vector field as repeated principal null
direction. The general case is when $\Phi$ has type $R_{15}$ and
the special case when $M$ has constant curvature is of this type
since then $ \dim \phi '_p  = 6$ for each $ p \in M$ . An example
of type $R_{7}$ can be obtained by adjusting the $R_{7}$ example
in the next section so that the 2-dimensional manifolds $M_{1}$
and $M_{2}$ described there have the same constant curvature. The
authors are not aware of a proper Einstein space of type $R_{14}$.

\section{Conformally Flat Space-times}
Here one has the decomposition (4) with the Weyl tensor
identically zero. Suppose $M$ admits a non-zero covariantly
constant vector field $k$. Then the Ricci identity on $k$ yields $
R_{abcd} k^d  = 0$ and hence $ R_{ab} k^b  = 0$ . A contraction of
(4) with $ k^a$ and a further contraction with $ k^c$ yields, in
turn, the relations
\begin{equation}
\left( {\rm i} \right) \ k_{[c} R_{d]b}  = {\textstyle{1 \over
3}}Rk_{[c} g_{d]b} \hspace{10mm \left( {\rm ii} \right) } \left(
{k_c k^c } \right)\left( {{\textstyle{1 \over 3}}Rg_{ab}  - R_{ab}
} \right) = {\textstyle{1 \over 3}}Rk_a k_b
\end{equation}
Now if $M$ is flat the holonomy type is $R_{1}$. Otherwise there
exists $ p \in M$ such that the curvature tensor (and hence the
Ricci tensor since $M$ is conformally flat) is not zero at $p$.
Thus if $M$ is not flat, (11)(ii) shows that $k$ is null at $p$ if
and only if $ R = 0$ at $p$. The same equation then shows that,
since $k$ is covariantly constant and nowhere zero, $k$ is null on
$M$ if and only if $ R \equiv 0$ on $M$.

Now if $k$ is null (11)(i) shows that $ R_{ab}  = \alpha k_a k_b$
on $M$ where $ \alpha :M \to {\bf R}$ (so the Ricci tensor, if not
zero, has Segre type $ \left\{ {\left( {211} \right)} \right\}$
with eigenvalue zero). But then the curvature tensor is
identically equal to $ E_{abcd}$ (since $ R \equiv 0$ on $M$) and
so, either from (4) after a short calculation or directly from the
classification in [17], one sees that the curvature is either zero
or constructed from a dual pair of null bivectors with principal
null direction $k$ at each point of $M$. A similar argument to
that in the vacuum case using the Ambrose-Singer theorem then
shows that $\Phi$ is of type $R_{8}$.

If $k$ is not null then at some $ p \in M$ $ R\left( p \right) \ne
0$ and (11)(ii) gives
\begin{equation}
R_{ab}  = {\textstyle{1 \over 3}}R\left( {g_{ab}  - \varepsilon
k_a k_b } \right)
\end{equation}
where $k$ has been assumed globally scaled so that $ k_a k^a  =
\varepsilon  =  \pm 1$ . From (12) and (4) one finds
\begin{equation}
R_{abcd}  = {\textstyle{1 \over 3}}R\left( {G_{abcd}  -
\varepsilon g_{a[c} k_{d]} k_b  - \varepsilon k_a k_{[c} g_{d]b} }
\right)
\end{equation}
Now if F is any bivector at $p$ satisfying $ F_{ab} k^b  = 0$ (13)
shows that
\begin{equation}
R_{abcd} F^{cd}  = {\textstyle{1 \over 3}}RF_{ab}  \ne 0
\end{equation}

But such bivectors form a 3-dimensional subspace $W$ of bivector
space at $p$ and the complementary subspace $ W^ *$ to $W$ is
spanned by three independent simple bivectors whose blades contain
$k$. Thus if $ H \in W^ *$ , $ R_{abcd} H^{cd}  = 0$ (since $
R_{abcd} k^d  = 0$ ). It follows that, since $ R\left( p \right)
\ne 0$ , the curvature tensor has rank equal to 3 at $p$. This
shows that $\phi$ has dimension $ 3$ and hence from table 1 that $
\dim \phi  = 3$ and that $\Phi$ is of type $R_{10}$ if $k$ is
spacelike and of type $R_{13}$ if $k$ is timelike. It follows from
(12) that $ R_{ab} k^b  = 0$ and that the Ricci and
energy-momentum tensors have Segre type $ \left\{ {1,(111)}
\right\}$ ($k$ timelike) and $ \left\{ {\left( {1,11} \right)1}
\right\}$ ($k$ spacelike).

Now suppose $M$ admits a recurrent (null) vector field $k$ such
that $ k_{a;b}  = k_a p_b$ . The Ricci identity on $k$ followed by
a contraction with $ g^{bd}$ gives
\begin{equation}
 k_a R^a _{\ bcd}  = 2k_b p_{[c;d]}  \hspace{5mm} \left( { \Rightarrow
 \ \ R_{ab} k^b  = 2p_{[a;b]} k^b } \right)
\end{equation}
The first of these shows that the bivector $ p_{[a;b]}$ at any $ p
\in M$ is amongst the bivectors contributed to $ \phi '_p$ , and
hence to $\phi$, by the curvature tensor as in (1). Thus from
theorem 1, it must have $k$ as an eigenvector and so $ p_{[a;b]}
k^b  = \nu k_a$ for $ \nu :M \to {\bf R}$ (or, alternatively, use
the identity $ R_{a[bcd]}  = 0$ to see that $ p_{[a;b]}$ is a
simple bivector whose blade contains $k$). Then a contraction of
(4) with $ k^a k^c$ shows that $ R \equiv 0$ on $M$. Thus $
R_{abcd}  = E_{abcd}$ on $M$ and then (8) and the argument in [2]
shows that $ \phi$ is dual invariant (as described earlier) and
that $ \dim \phi$ is even. Since $k$ may be assumed here to be not
covariantly constant, it follows that the only possibilities for
$\Phi$ are the types $R_{7}$ and $R_{14}$. The following theorem
has been proved (c.f. [3])

\begin{theorem}
Let $M$ be a simply connected conformally flat, not flat
space-time. Then the holonomy group $\Phi$ of $M$ is one of the
types $R_{7}$, $R_{8}$, $R_{10}$, $R_{13}$, $R_{14}$ or $R_{15}$.
\end{theorem}

It is clear from the proof that, given the holonomy group is
reducible, the condition that $ R \equiv 0$ on $M$ is equivalent
to $M$ admitting a null recurrent (including covariantly constant)
vector field and then $\Phi$ is one of the types $R_{7}$, $R_{8}$
or $R_{14}$. If a recurrent (not globally scalable to be
covariantly constant, and hence null) vector field is admitted
then the possibilities for $\Phi$ are the types $R_{7}$ and
$R_{14}$.

A closer investigation of the curvature tensor and the tensor $
E_{abcd}$ reveals further information about the Segre type of the
Ricci tensor when the holonomy group is reducible [3]. In fact in
such cases this Segre type at $ p \in M$ (which may vary with $p$)
is, for a non-zero Ricci tensor, either

\begin{enumerate}
\item $ \left\{ {\left( {31} \right)} \right\} $ with eigenvalue
zero $ \left( { \Rightarrow R = 0} \right) $ \item $ \left\{
{2\left( {11} \right)} \right\} $ with eigenvalues differing only
in sign $ \left( { \Rightarrow R = 0} \right) $ \item $ \left\{
{\left( {211} \right)} \right\} $ with eigenvalue zero $ \left( {
\Rightarrow R = 0} \right) $ \item $ \left\{ {1,\left( {111}
\right)} \right\} $ with the non-degenerate (timelike) eigenvalue
zero $ \left( { \Rightarrow R \ne 0} \right) $ \item $ \left\{
{\left( {1,11} \right)1} \right\} $ with the non-degenerate
eigenvalue zero $ \left( { \Rightarrow R \ne 0} \right) $ \item $
\left\{ {\left( {1,1} \right)\left( {11} \right)} \right\}  $ with
eigenvalues differing only in sign $ \left( { \Rightarrow R = 0}
\right) $
\end{enumerate}

Examples of each of these holonomy type possibilities, except one,
can now be given. (The authors are unaware of an example of the
$R_{14}$ case). For the $R_{7}$ type let $ \left( {M_1 ,g_1 }
\right)$ and $ \left( {M_2 ,g_2 } \right)$ be two 2-dimensional
manifolds with $ M_1  = M_2  = {\bf R}^2$ and $ g_1$ and $ g_2$
metrics on $ M_1$ and $ M_2$ with signatures $ \left( { - 1,1}
\right)$ and $ \left( {1,1} \right)$ respectively. Suppose also
that $ \left( {M_1 ,g_1 } \right)$ has constant curvature $ k < 0$
and that that $ \left( {M_2 ,g_2 } \right)$ has constant curvature
$ - k$ . Then $ M = M_1  \times M_2$ is a connected and
simply-connected space-time with Lorentz metric $ g = g_1  \times
g_2$ . Further if $ x^0 ,x^1$ are coordinates for $ M_1$ and $ x^2
,x^3$ are coordinates for $ M_2$ then the metric $ g$ on $ M$ is
represented in the coordinate system $ x^a$ on $ M$ by
\[ ds^2  =
g_{\alpha \beta } dx^\alpha  dx^\beta   + g_{AB} dx^A dx^B
\]
where $ \alpha ,\beta ,\gamma  = 0,1$ , $ A,B,C = 2,3$ and where $
g_{\alpha \beta }  = g_{\alpha \beta } \left( {x^\gamma  } \right)
$ and $ g_{AB}  = g_{AB} \left( {x^C } \right)$ are the components
of $ g_1$ and $ g_2$ in the coordinates $ x^\alpha$ and $ x^A$,
respectively. For $ p \in M$ one can construct a real null tetrad
$ \left( {l,n,x,y} \right)$ in some coordinate neighbourhood $U$
of $p$ such that $l$ and $n$ are tangent to $ M_1$ and $x$ and $y$
are tangent to $ M_2$ in $U$. Then the curvature tensor in $U$ can
be written in terms of the bivector $ A = l \wedge n$ and its dual
$ \mathop A\limits^ *   = x \wedge y$ as
\[ R_{abcd}  = a\left(
{A_{ab} A_{cd}  + \mathop {A_{ab} }\limits^ *  \mathop {A_{cd}
}\limits^ * } \right) \hspace{5mm} \left( {0 \ne a = {\rm
constant}} \right)
\]
It follows that the Ricci scalar $ R = 0$ and, from (4), that the
Weyl tensor is zero. The Ricci tensor has Segre type $ \left\{
{\left( {1,1} \right)\left( {11} \right)} \right\}$ with
eigenvalues constant and differing only in sign and $ \left( {M,g}
\right)$ represents a conformally flat non-null Einstein-Maxwell
field of the Bertotti-Robinson type [4]. Also $l$ and $n$ may be
extended to null vector fields on $M$ which are recurrent (but
which cannot be scaled so as to be covariantly constant) and the
bivectors $ A$ and $ \mathop A\limits^ *$ are covariantly
constant. It follows that the holonomy type of $ \left( {M,g}
\right)$ is $R_{7}$ (see [7]).

An example of a type $R_{8}$ holonomy group is given by the plane
wave space-time given in a global coordinate system $ \left(
{x,y,u,v} \right)$ on $ {\bf R}^4$ by the metric [4]
\[
ds^2  = dx^2  + dy^2  - 2dudv - h\left( u \right)\left( {x^2  +
y^2 } \right)du^2
\]
where $ h:{\bf R}^4  \to {\bf R}$ is everywhere positive. This
space-time is conformally flat and represents a null Einstein-
Maxwell field. The covector field $ l_a  = u,_a$ is global, null
and covariantly constant as are the bivectors $ F_{ab}  = 2l_{[a}
x_{,b]}$ and $ \mathop {F_{ab} }\limits^ *   =  2l_{[a} y_{,b]}$ .
The curvature tensor is
\[
R_{abcd}  = f\left( u \right)\left( {F_{ab} F_{cd}  + \mathop
{F_{ab} }\limits^ *  \mathop {F_{cd} }\limits^ *  } \right)
\]
for $ f:{\bf R}^4  \to {\bf R}$ positive.

Next let $ M' = {\bf R}^3$ and $ g'$ be a Lorentz metric on $ M' $
which is of non-zero constant curvature. Then the usual metric
product $ M = M' \times {\bf R} $ yields a space-time admitting a
global covariantly constant spacelike vector field. The holonomy
group is easily checked to be of type $R_{10}$ (c.f. [7]) and
since a 7-dimensional Lie algebra of (local) Killing vector fields
is admitted by $M$, it is conformally flat.

A simply connected region of the well-known Einstein static
universe which admits a global covariantly constant timelike
vector field is a conformally flat perfect fluid and an example of
holonomy type $R_{13}$. The standard (simply connected)
Friedmann-Robertson-Walker space-times are conformally flat and
give examples of holonomy type $R_{15}$.

\section{Null Einstein-Maxwell Fields}
In this case let $F$ be the nowhere zero global null Maxwell
bivector with principal null direction represented by the nowhere
zero global null vector field $l$. Then in a coordinate
neighbourhood of some (any) $ p \in M$ there exists a (nowhere
zero) spacelike vector field $q$ orthogonal to $l$ such that $
F_{ab}  = 2l_{[a} q_{b]}$ and
\begin{equation}
R_{abcd}  = C_{abcd}  + {\textstyle{\kappa  \over 2}}\left(
{F_{ab} F_{cd}  + \mathop {F_{ab} }\limits^ *  \mathop {F_{cd}
}\limits^ *  } \right)
\end{equation}
where the Einstein-Maxwell equations are $ R_{ab}  = \mu l_a l_b$
$ \left( { \Rightarrow R = 0} \right)$ and where $ \mu :M \to {\bf
R}$ is nowhere zero.

Suppose first that $\Phi$ is reducible and, as before, let $k$ be
a recurrent vector field on $M$. If $k$ is covariantly constant
then the Ricci identity gives $ R_{abcd} k^d  = 0$ , $ R_{ab} k^b
= 0$ and hence $ k_a l^a  = 0$ . Thus $k$ is everywhere null or
everywhere spacelike.

If $k$ is everywhere null one may take $ l = k$ and so from (16) $
C_{abcd} l^d  = 0$ . Thus the Petrov type at any $ p \in M$ is $N$
(with the repeated principal null direction spanned by $l$) or
$O$. If $M$ is conformally flat, (16) together with the
Ambrose-Singer theorem and the covariant constancy of $l$ shows
that $\Phi$ is of type $R_{8}$ (c.f. section 5). If the Petrov
type is $N$ at some $ p \in M$ then the Weyl tensor is also
constructed out of the bivectors F and $ \mathop F\limits^ *$ at
$p$ and a further use of the Ambrose-Singer theorem and (16) shows
that $\phi$ contains either just the two independent bivectors $F$
and $ \mathop F\limits^ *$ or (if judicious cancellation occurs)
just one of these bivectors. In the former case $\Phi$ is of type
$R_{8}$ and in the latter case of type $R_{3}$.

If $k$ is spacelike one may choose a real null tetrad $ \left(
{l,n,x,y} \right)$ in a coordinate neighbourhood of some (any) $ p
\in M$ such that $ k = x$ and so $ R_{abcd} x^d  = 0$ . It follows
that the curvature tensor is constructed in the neighbourhood from
the bivectors $ l \wedge n$, $ l \wedge y$ and $ n \wedge y$ and
for consistency with the expression for the Ricci tensor, one must
then have in this neighbourhood
\begin{equation}
R_{abcd}  = 4\mu l_{[a} y_{b]} l_{[c} y_{d]}
\end{equation}
The Ambrose-Singer theorem then shows that $\phi$ is spanned by
null bivectors which contract to zero with $x$ (since $x$ is
covariantly constant). From table 1 this can easily be seen to
eliminate the type $R_{6}$ but not the types $R_{3}$ and $R_{10}$.
(Alternatively if one assumes $\Phi$ is of type $R_{6}$ then $l$
would be recurrent and the Ambrose-Singer theorem would then show
that $\phi$ consisted only of $ l \wedge y$ and give the
contradiction that the type was $R_{3}$). The only possibilities
are that $\Phi$ is of type $R_{3}$ or type $R_{10}$. If the type
is $R_{3}$ then $l$ is covariantly constant on $M$ (see previous
paragraph). It also follows from (16) and (17) that for either
holonomy type the Petrov type is $N$ at every point of $M$ with
repeated principal null direction spanned by $l$.

Now suppose that $k$ is a recurrent vector field on $M$ but that
$M$ admits no covariantly constant non-zero vector fields (so that
$k$ is null on $M$). Then $ k_a R^a _{\ bcd}  = 2k_b p_{[c;d]}$
and by the argument given in the previous section $ 2p_{[a;b]} k^b
= \lambda k_a$ for some $ \lambda :M \to {\bf R}$ . It follows
that $ \lambda k_a  = \mu \left( {l_c k^c } \right)l_a$ and hence
that, irrespective of the possible zeros of $\lambda$, $l$ and $k$
are proportional at each $ p \in M$ . Thus one may take $ k = l$
and then (16) and the Ricci identity for $l$ give
\begin{equation}
 l_a R^a _{\ bcd}  =
2l_b p_{[c;d]}  = l_a C^a _{bcd}  \hspace{5mm} \left( {
\Rightarrow l_a l_{[e} C^a _{b]cd}  = 0} \right)
\end{equation}
This shows that the Petrov type at any $ p \in M$ is $III$, $N$ or
$O$ with $l$ spanning the repeated null direction in the first two
cases. If the Petrov type is nowhere type $III$ then (16) and the
Ambrose-Singer theorem and theorem 1 yield the contradiction that
$l$ is covariantly constant on $M$ and so the Petrov type must be
$III$ on some (necessarily open because of the remarks on rank in
section 2) subset of $M$. From the algebraic structure of a type
$III$ Weyl tensor at $ p \in M$ one can extend $l$ to a real null
tetrad $ \left( {l,n,x,y} \right)$ at $p$ such that, using (16),
the simple bivectors $ l \wedge n$ , $ x \wedge y$ and $ l \wedge
x$ (at least) are contributed to $ \phi '_p$ and hence to $\phi$
through (1). Thus $ \dim \phi\geq 3$ and the conditions on $l$ and
table 1 limit the possibilities for $\Phi$ to the types $R_{9}$,
$R_{12}$ and $R_{14}$. But the case $R_{9}$ requires each member
of $\phi$ to be simple and with $l$ in its blade whilst it is
easily checked that if $\Phi$ is of type $R_{12}$ then $\phi$ does
not admit three independent {\it {simple}} bivectors. Hence $\Phi$
is of type $R_{14}$. One can thus state the following theorem.

\begin{theorem}
Let $M$ be a simply connected space-time of the null
Einstein-Maxwell type. Then the holonomy group $\Phi$ of $M$ is of
the type $R_{3}$, $R_{8}$, $R_{10}$, $R_{14}$ or $R_{15}$.
\end{theorem}

It is remarked that, whatever the holonomy type, the Petrov type
is algebraically special (including type $O$) at any $ p \in M$
with $l$ spanning a repeated principal null direction, by virtue
of the Goldberg-Sachs theorem [18].

Examples of null Einstein-Maxwell fields of holonomy type $R_{3}$
can be found in [9] and [19] and examples of types $R_{8}$,
$R_{10}$ and $R_{14}$ are also displayed in [9]. Since the
reducible possibilities described in theorem 5 were found at each
point to be of Petrov type $O$, $N$ or $III$, any type $D$ null
Einstein-Maxwell fields (which exist - see [4], p256) are of
holonomy type $R_{15}$.

\section{Non-Null Einstein-Maxwell Fields}
Here the global Maxwell bivector is a non-null bivector F. One
assumes the existence of independent global null vector fields $l$
and $n$ and functions $\alpha$ and $\beta$ such that $ \alpha ^2 +
\beta ^2$ is nowhere zero on $M$ and such that in local
coordinates
\begin{equation} F_{ab}  = 2\alpha l_{[a} n_{b]}  +
2\beta x_{[a} y_{b]}
\end{equation}
The Einstein-Maxwell equations then give
\begin{equation}
R_{ab}  =  - {\textstyle{\kappa  \over 2}}\left( {\alpha ^2  +
\beta ^2 } \right)\left( {2l_{(a} n_{b)}  - x_a x_b  - y_a y_b }
\right)
\end{equation}
which has Segre type $ \left\{ {\left( {1,1} \right)\left( {11}
\right)} \right\}$ with its two distinct eigenvalues differing
only in sign and nowhere zero on $M$.

Suppose that $M$ admits a recurrent vector field $k$. Clearly $k$
cannot be covariantly constant since then a contraction of the
Ricci identity for $k$ reveals a zero eigenvalue of the Ricci
tensor at each $ p \in M$ which contradicts (20). Thus $k$ can be
assumed null and the Ricci identity for $k$ is then $ k_a R^a
_{bcd}  = 2k_b p_{[c;d]}$ and so, as before, the bivector $
p_{[a;b]}$ is contained in $\phi$ and has $k$ as an eigenvector at
each $ p \in M$ . A contraction of this Ricci identity then shows
that $k$ everywhere spans a null Ricci eigendirection. But (20)
shows that at each $ p \in M $ the only null Ricci eigendirections
are spanned by $l$ and $n$. So one can assume, say, that $ k = l$
on $M$. Theorem 1 then shows that the bivectors spanning $\phi$
are (some subset of) the bivectors $ A_{ab}  = 2l_{[a} n_{b]}$ , $
B_{ab}  = 2l_{[a} x_{b]}$ and their duals. Thus the curvature
tensor must be constructed only from these bivectors and, in
addition, must be consistent with (20). It follows that the local
expression for the curvature tensor is
\begin{eqnarray} R_{abcd} = & {\textstyle{\kappa  \over 2}}\left( {\alpha ^2 +
\beta ^2 } \right)\left( {A_{ab} A_{cd}  + \mathop {A_{ab}
}\limits^ * \mathop {A_{cd} }\limits^ *  } \right) \nonumber \\ &
+ a\left({B_{ab} B_{cd}  - \mathop {B_{ab} }\limits^ *  \mathop
{B_{cd} }\limits^ *  } \right) + b\left( {B_{ab} \mathop {B_{cd}
}\limits^*   + \mathop {B_{ab} }\limits^ *  B_{cd} } \right)
\nonumber \\ & + c\left( {A_{ab} B_{cd}  + B_{ab} A_{cd}  -
\mathop {A_{ab} }\limits^ * \mathop {B_{cd} }\limits^ *   -
\mathop {B_{ab} }\limits^ * \mathop {A_{cd} }\limits^ *  } \right)
\\ & + d\left( {A_{ab} \mathop {B_{cd} }\limits^ *   + \mathop
{B_{ab} }\limits^
*  A_{cd}  + B_{ab} \mathop {A_{cd} }\limits^ *   + \mathop
{A_{ab} }\limits^ *  B_{cd} } \right) \nonumber
\end{eqnarray}
for functions $a$, $b$, $c$ and $d$ . Since $ \alpha ^2  + \beta
^2$ is nowhere zero it can be shown from (21) and some elementary
algebra that the curvature tensor contributes at least two
independent bivectors to $\phi$ (through (1)) and that one of them
is (simple) spacelike and is contracted to zero by $l$. Since no
covariantly constant vector fields are admitted the possibilities
for $\Phi$ are $R_{7}$, $R_{9}$, $R_{12}$ and $R_{14}$. However
the algebras for $R_{9}$ and $R_{12}$ cannot admit a spacelike
bivector with the above properties (because all members of $R_{9}$
are simple with their blades containing $l$ and the only simple
members of $R_{12}$ are linear combinations of $ l \wedge x$ and $
l \wedge y$ and hence null). So only $R_{7}$ and $R_{14}$ remain.
The following result has been established.

\begin{theorem}
Let $M$ be a simply connected space-time of the non-null
Einstein-Maxwell type. Then the holonomy group $\Phi$ of $M$ is of
the type $R_{7}$, $R_{14}$ or $R_{15}$.
\end{theorem}

The Ricci identity above for the recurrent null vector field $k$
and the fact that $k$ is an eigenvector of $ p_{[a;b]}$ together
show that
\begin{equation}
R_{abcd} k^a k^c  = ek_b k_d  \hspace{5mm} R_{ab} k^b  = fk_a
\end{equation}
for functions $e$ and $f$ on $M$. These, in turn, lead from (4)
(or [17]) to $ C_{abcd} k^a k^c  \propto k_b k_b$. Hence, at
points where the Weyl tensor is not zero, it is algebraically
special with $k$ spanning a repeated principal null direction. If
the holonomy type is $R_{7}$ two such recurrent null vector fields
are admitted and the Petrov type at any $ p \in M$ is thus $O$ or
$D$. For the $R_{14}$ type only one such recurrent null vector
field occurs and so at any $ p \in M$ the Petrov type could be any
of the algebraically special types (including $O$).

An example of a non-null (conformally flat) Einstein-Maxwell field
of holonomy type $R_{7}$ was given in section 5. An example of the
holonomy type $R_{14}$ is listed in [9]. A (simply connected)
Reissner-Nordstrøm space-time is an example of holonomy type
$R_{15}$.

\section{Perfect Fluid Space-Times}
Here one assumes a global, nowhere zero, unit timelike vector
field $u$ (the fluid flow vector) and functions $p$ (the pressure)
and $\mu$ (the energy density) on $M$. It is assumed that $\mu$
and $p$ do not simultaneously vanish over some non-empty open
subset of $M$. One then has an energy- momentum tensor with
components $T_{ab}$ given by
\begin{eqnarray}
T_{ab}  = \left( {\mu  + p} \right)u_a u_b  + pg_{ab} \\
 {\left( \Rightarrow R_{ab}  = \kappa \left[ {{\textstyle{1 \over
2}}\left( {\mu  - p} \right)g_{ab}  + \left( {\mu  + p} \right)u_a
u_b } \right]\right)} \nonumber
\end{eqnarray}
Suppose that $k$ is a recurrent (possibly covariantly constant)
vector field on $M$. Then, by a similar argument to that given in
the previous cases, $k$ is a Ricci eigenvector, $ R_{ab} k^b  =
\lambda k_a$ for some function on $M$, and so from (23)
\begin{equation}
\lambda k_a  = {\textstyle{\kappa  \over 2}}\left( {\mu  - p}
\right)k_a  + \kappa \left( {\mu  + p} \right)\left( {k^b u_b }
\right)u_a
\end{equation}
Now since $k$ and $u$ are nowhere zero it follows that if $k$ is
null $ \left( {k^b u_b } \right)$ is nowhere zero and so $ \left(
{\mu  + p} \right) = 0$ everywhere on $M$. Thus, from (23), $M$ is
an Einstein space(-time) with $ \mu  - p$ constant on $M$. If $
\mu  - p = 0$  $M$ is vacuum and the conclusions of section 3 hold
whereas if $ \mu  - p \ne 0$ section 4 completes the argument.

If $k$ is timelike (and hence one can assume $k$ is covariantly
constant) then (24) with $ \lambda  \equiv 0$ and the facts that
$u$ and $k$ are nowhere zero on $M$ show that at each $ p \in M$
$\mu  + p = 0$ if and only if $ \mu  - p = 0$ . Recalling the
exclusion clause at the beginning of this section one sees that $
\mu  - p$ and $ \mu  + p$ cannot both vanish over any non-empty
open subset of $M$ and so one may take $ k = u$ on $M$. Thus $u$
is covariantly constant on $M$ and the Ricci identity shows that $
R_{abcd} u^d  = 0\left( { \Rightarrow R_{ab} u^b  = 0} \right)$
and so $ 3p + \mu  = 0$ on $M$ and
\begin{equation}
R_{ab}  =  -2\kappa p\left( {g_{ab}  + u_a u_b } \right)
\end{equation}
The contracted Bianchi identity then shows that $p$ and hence
$\mu$ and $p$ separately are (non-zero) constants on $M$. If one
introduces a pseudo-orthonormal tetrad $ \left( {u,x,y,z} \right)
$ at any $ p \in M$ one finds that, for consistency with (25), the
curvature tensor takes the form
\begin{equation}
R_{abcd}  =  - \kappa p\left( {X_{ab} X_{cd}  + Y_{ab} Y_{cd}  +
Z_{ab} Z_{cd} } \right)
\end{equation}
where $ X_{ab}  = 2x_{[a} y_{b]}$ , $ Y_{ab}  = 2x_{[a} z_{b]}$
and $ Z_{ab}  = 2y_{[a} z_{b]}$ . Clearly, the holonomy group
$\Phi$ is of type $R_{13}$.

If $k$ is spacelike (and again assumed covariantly constant) then
(24) with $ \lambda  \equiv 0$ shows that $ k_a u^a  = 0$ over an
open dense subset of $M$ and hence on $M$ and then that $ \mu  =
p$ on $M$. Thus
\begin{equation}
 R_{ab}  = 2\kappa pu_a u_b
 \end{equation}
A similar expression to (26) can be found for the curvature tensor
at $ p \in M$ in terms of the pseudo-orthonormal tetrad given in
an obvious notation by $ \left( {u,k,y,z} \right)$

\begin{equation}
R_{abcd}  = \kappa p\left( {C_{ab} C_{cd}  + D_{ab} D_{cd}  +
E_{ab} E_{cd} } \right)
\end{equation}
where $ C_{ab}  = 2u_{[a} y_{b]}$ , $ D_{ab}  = 2u_{[a} z_{b]}$
and $ E_{ab}  = 2y_{[a} z_{b]}$ . Clearly, the holonomy group
$\Phi$ is of type $R_{10}$. Thus it is not possible for both a
timelike and a spacelike covariantly constant vector field to be
admitted. When $k$ is timelike it is clear from (25) and (26) that
the curvature tensor and the Ricci tensor tetrad components are
the same for each of a 3-parameter family of pseudo-orthonormal
tetrads (that is they are unchanged under the 3-dimensional
subgroup $SO$(3) of the Lorentz group. Hence, from (4), the Weyl
tensor is similarly unchanged. Since the largest such subgroup for
a non-zero Weyl tensor has dimension 2 [14] it follows that in
this case, $M$ is conformally flat. The following theorem is
established.

\begin{theorem}
Let $M$ be a simply connected space-time representing a perfect
fluid. Then if $M$ is not an Einstein space(-time) the holonomy
group $\Phi$ of $M$ is of type $R_{10}$, $R_{13}$ or $R_{15}$.
\end{theorem}

The Einstein static universe discussed in section 5 is an example
with a type $R_{13}$ holonomy group whilst the Gödel metric
provides an example of type $R_{10}$. The standard Friedmann-
Robertson-Walker metrics give examples of type $R_{15}$.

\section{Massive Scalar Fields}
If $M$ is the space-time of a massive scalar field then one has a
scalar field $\psi$ defined on $M$ such that $\psi$ is not
constant over any non-empty open subset of $M$ together with a
non-zero constant m such that the associated energy-momentum
tensor is given by [20]
\begin{equation}
T_{ab}  = \psi _a \psi _b  - {\textstyle{1 \over 2}}\left( {\psi
^c \psi _c  + m^2 \psi ^2 } \right)g_{ab}
\end{equation}
where $ \psi _a  = \psi _{,a}$ . Then (29) and the contracted
Bianchi identity give
\begin{equation}
R_{ab}  = \kappa \psi _a \psi _b  + {\textstyle{1 \over 2}}\kappa
m^2 \psi ^2 g_{ab}  \hspace{10mm} \psi _{,a}^a  = m^2 \psi
\end{equation}
Now (30) reveals that at any $ p \in M$ , any $ v \in T_p M$
orthogonal to $ \psi _a$ is a Ricci eigenvector with the same
eigenvalue $ {\textstyle{1 \over 2}}\kappa m^2 \psi ^2$ . Hence
the Ricci tensor has a triple (real) eigenvalue degeneracy and so
has Segre type either $ \left\{ {\left( {211} \right)} \right\}$ ,
$ \left\{ {\left( {1,11} \right)1} \right\}$ or $ \left\{
{1,\left( {111} \right)} \right\}$ at $p$ [17].

Now suppose $M$ admits a recurrent vector field $k$. Then, by an
argument identical to one used several times earlier, the Ricci
identity on $k$ shows that $k$ is a Ricci eigenvector everywhere
on $M$. If $k$ is null then one can write down, using the argument
of the previous section, an expression for the curvature tensor
similar to (21) but now one must impose consistency with the first
equation in (30). Since $k$ is a null Ricci eigenvector the Segre
type of the Ricci tensor at any $ p \in M $ must be $
\left\{{\left( {211} \right)} \right\}$ or $ \left\{ {\left(
{1,11} \right)1} \right\}$ and this expression for the curvature
tensor, on contraction, shows that the latter is impossible. Thus
the Segre type is everywhere $ \left\{ {\left( {211} \right)}
\right\}$ . But the first equation in (30) cannot be of this type
unless $ \psi _a$ spans the null Ricci eigendirection at those
points where $ \psi _a$ is not zero. So by the uniqueness of the
null Ricci eigendirection for this type it follows that $ \psi _a
= \sigma k_a$ for some function $\sigma$ on $M$. Hence, since $
k_{a;b} = k_a p_b$ and $ \psi _{a;b}  = \psi _{b;a}$ one finds
that $ \psi _{a;b}  = \nu \psi _a \psi _b$ for some function $\nu$
on $M$. The second equation in (30) then gives the contradiction
that $ \psi  \equiv 0$ on $M$.

If $k$ is not null (and assumed globally scaled so that $ k_{a;b}
= 0$ and $ k^a k_a  = \varepsilon  =  \pm 1$ ) the Ricci identity
gives $ R_{ab} k^b  = 0$ everywhere on $M$. This fact and the
first equation of (30) yields

\begin{equation}
\left( {\psi _b k^b } \right)\psi _a  + {\textstyle{1 \over 2}}m^2
\psi ^2 k_a  = 0
\end{equation}
It follows that $ \psi _a k^a$ may only vanish over a subset of
$M$ with no interior and so over the complement of this subset,
and hence over $M$, $ \psi _a  = \rho k_a$ for some function $
\rho :M \to {\bf R}$ . The previous equation then gives
\begin{equation}
\varepsilon \rho ^2  + {\textstyle{1 \over 2}}m^2 \psi ^2  = 0
\end{equation}
Now in some open neighbourhood $U$ of any $ p \in M$ one may write
$ k_a  = u_{,a}$ for some function $u$ on $U$. It then follows
(since $ k_{a;b}  = 0 \Rightarrow \rho  = \rho \left( u \right)$ )
that on $U$, $ \psi _{a;b}  = \dot \rho k_a k_b$ (where a dot
represents differentiation with respect to $u$). Also $ \psi  =
\psi \left( u \right)$ and $ \dot \psi  = \rho$ . The second
equation in (30) then reveals that $ \varepsilon \dot \rho  = m^2
\psi$ on $U$. But on differentiating (32) one finds that $
2\varepsilon \rho \dot \rho  + m^2 \psi \dot \psi  = 0$ on $U$.
Thus $ \rho  = 0$ and the contradiction $ m^2 \psi  = 0$ on $M$ is
achieved. This gives the following result.

\begin{theorem}
Let $M$ be a simply connected space-time representing a massive
scalar field. Then the holonomy group $\Phi$ of $M$ is of the type
$R_{15}$.
\end{theorem}

\section{Conclusions}
The holonomy group structure of simply connected space-times which
have certain specified energy-momentum tensor (or which are
conformally flat) has been studied. The allowed holonomy types are
(excluding the flat space-time $R_{1}$ type) for vacuum
space-times (types $R_{8}$, $R_{14}$, $R_{15}$), for Einstein
space(-times) with $ R \ne 0$ ($R_{7}$, $R_{14}$, $R_{15}$), for
conformally flat space-times ($R_{7}$, $R_{8}$, $R_{10}$,
$R_{13}$, $R_{14}$, $R_{15}$), for null Einstein-Maxwell
space-times ($R_{3}$, $R_{8}$, $R_{10}$, $R_{14}$, $R_{15}$),  for
non-null Einstein-Maxwell space-times ($R_{7}$, $R_{14}$,
$R_{15}$), for perfect fluid space-times  ($R_{10}$, $R_{13}$,
$R_{15}$) and for massive scalar fields ($R_{15}$). Some of the
above reducible cases (i.e. not type $R_{15}$) are, in fact,
non-degenerately reducible (see caption to table 1). These space-
times are locally the metric product of lower dimensional
manifolds and so allow a natural product coordinate system
convenient for their exploration. Similar (but not always quite so
convenient) coordinate systems are admitted in the degenerately
reducible cases.

Some results which arise are perhaps worth general statements. If
$M$ admits a null recurrent (not necessarily covariantly constant)
vector field $k$ then $M$ is algebraically special (including type
$O$) in the Petrov classification and $k$ spans a repeated
principal null direction of the Weyl tensor and a null Ricci
eigendirection. Suppose, on the other hand that $M$ admits a
non-null covariantly constant vector field $k$. Then if $k$ is
timelike the Ricci identity shows that $k$ is a timelike Ricci
eigenvector and so the Ricci tensor is of the diagonalisable type
(i.e. Segre type $ \left\{ {1,111} \right\}$ or one of its
degeneracies (see e.g.[17]). It is then easily checked that, in
the usual $6\times6$ formulation, the tensors $ R_{abcd}$ and $
E_{abcd} $ are (in an obvious algebraic sense) diagonalisable and
hence so is the Weyl tensor. This last statement means that the
Petrov type is $O$, $I$ or $D$. If a covariantly constant
spacelike vector field is admitted then, potentially, any Petrov
type might occur (see [7] and c.f. [21]).

\subsection*{Acknowledgements} One of the authors (G.S.H.)
acknowledges with thanks the award of a N.A.T.O. grant (number CRG
CRG 960140).


\end{document}